\documentclass[10pt,journal,compsoc]{IEEEtran}
\ifCLASSOPTIONcompsoc
  \usepackage[nocompress]{cite}
\else
  \usepackage{cite}
\fi
\ifCLASSINFOpdf
  \usepackage[pdftex]{graphicx}
  \graphicspath{{./Images/}}
  \DeclareGraphicsExtensions{.pdf,.jpeg,.png,.PNG}
\else
\fi
\usepackage{amsmath}
\usepackage{algorithmic}

\usepackage{array}

\ifCLASSOPTIONcompsoc
  \usepackage[caption=false,font=footnotesize,labelfont=sf,textfont=sf]{subfig}
\else
 \usepackage[caption=false,font=footnotesize]{subfig}
\fi
 \let\MYorigsubfloat\subfloat
 \renewcommand{\subfloat}[2][\relax]{\MYorigsubfloat[]{#2}}
\usepackage{url}

\usepackage[justification=centering]{caption}
\usepackage{makecell}
\usepackage{multirow}
\usepackage{multicol}

\begin{document}

\copyright 2021 IEEE. Personal use of this material is permitted. Permission from IEEE must be obtained for all other uses, in any current or future media, including reprinting/republishing this material for advertising or promotional purposes, creating new collective works, for resale or redistribution to servers or lists, or reuse of any copyrighted component of this work in other works.

Published version is available: 

\url{https://doi.org/10.1109/TITS.2021.3110949}

 \newpage

\title{The Devil Is in the Details: An Efficient Convolutional Neural Network for Transport Mode Detection}
%
%
%
%

\author{Hugues~Moreau, 
        Andrea~Vassilev, 
        and~Liming~Chen,~\IEEEmembership{Senior~IEEE~member}
\IEEEcompsocitemizethanks{
\IEEEcompsocthanksitem H. Moreau and A. Vassilev are with the Universite Grenoble Alpes, CEA, Leti, F-38000 Grenoble, France
; E-mail: \{hugues.moreau,~andrea.vassilev\}@cea.fr\protect

\IEEEcompsocthanksitem L. Chen is with the Computer science and mathematics department at École Centrale de Lyon, France.
E-mail: liming.chen@ec-lyon.fr \protect\\
}
\thanks{Manuscript received Sept, 2020; revised June, 2021.}}

%
%

\ifCLASSOPTIONpeerreview
    \markboth{Journal of \LaTeX\ Class Files,~Vol.~14, No.~8, August~2015}%
    {Moreau \MakeLowercase{\textit{et al.}}: An efficient Convolutional Neural Network for Transport Mode detection}
\fi
%

\IEEEpubid{\copyright~2021 IEEE}

\IEEEtitleabstractindextext{%
\begin{abstract}
Transport mode detection is a classification problem aiming to design an algorithm that can infer the transport mode of a user given multimodal signals (GPS and/or inertial sensors). It has many applications, such as carbon footprint tracking, mobility behaviour analysis, or real-time door-to-door smart planning. Most current approaches rely on a classification step using Machine Learning techniques, and, like in many other classification problems, deep learning approaches usually achieve better results than traditional machine learning ones using handcrafted features. Deep models, however, have a notable downside: they are usually heavy, both in terms of memory space and processing cost. We show that a small, optimized model can perform as well as a current deep model. During our experiments on the GeoLife and SHL 2018 datasets, we obtain models with tens of thousands of parameters, that is, 10 to 1,000 times less parameters and operations than networks from the state of the art, which still reach a comparable performance. We also show, using the aforementioned datasets, that the current preprocessing used to deal with signals of different lengths is suboptimal, and we provide better replacements. Finally, we introduce a way to use signals with different lengths with the lighter Convolutional neural networks, without using the heavier Recurrent Neural Networks.
\end{abstract}

}

\maketitle

\IEEEdisplaynontitleabstractindextext

%
\IEEEpeerreviewmaketitle

\IEEEraisesectionheading{\section{Introduction}}

\IEEEPARstart{T}{ransport} mode detection is a family of classification problems in which an algorithm has to predict the transport mode of a given user, using several signals, whether it is GPS or inertial sensors (or both). The exact list of possible transport modes may vary depending on the application, but most applications include at least the most common ones: Walk, Bike, or Car, for instance. The applications are numerous: real-time trajectory planner (one that avoids traffic jams to drivers, for instance); automatic carbon footprint estimator; or mobility behaviour analysis. 
The signals are collected from an embedded device (either the sensors of a mobile phone, or a dedicated device), and processed by a Transport Mode Detection Algorithm, in order to determine the transport mode of the owner of the device. This algorithm has multiple steps that often include cleaning, preprocessing, and classification in itself. The latter operation is the most diverse from one algorithm to the other. All algorithms use Machine Learning \textit{i.e.}, methods that use a certain amount of labeled data to learn how to predict the output transport mode, before making predictions on unseen samples. But the approaches can be classified into two families: the 'traditional' Machine Learning approach, relying on handcrafted features (as in \cite{zhu_learning_2016, zheng_understanding_2010, etemad_transportation_2018, dodge_revealing_2009, maenpaa_travel_2017, zheng_learning_2008}), and the Deep Learning one, in which the classification is learnt using neural networks (see \cite{liu_end--end_2017, yu_travel_2019, huang_densely_2017, endo_classifying_2016}).  
The traditional Machine Learning method involves using handcrafted features which are assumed to be relevant enough to solve the problem, while the Deep Learning approach replaces these handcrafted features by features that are automatically learnt by a deep neural network. This approach generally yields better performance than the traditional one, but at a cost: the memory and computation requirements are dramatically increased, thus limiting the development of embedded systems. Yet, embedded applications are a great field of application for neural networks \cite{wu_lstm_2020}. 

To understand the kind of constraints embedded applications bring, let us take an example with a practical use case. Let us consider a carbon footprint automatic estimator on a  smartphone. The end user would download an application, and record the beginning and end of their trips. The application automatically estimates the transport mode of the user to compute an estimation of the greenhouse gas the trip emitted.  Compared to a client-server application, an embedded classifier allows to guarantee privacy to the users, and can function despite uncertain wireless data coverage. However, if this embedded classifier is not resource-efficient, the application will end up draining the battery of the end users, who might choose not to use it. Hence, resource efficiency is key for embedded applications.

When dealing with embedded devices, most neural networks are trained offline (on a computer) on prerecorded data, and sent to embedded devices for inference. This method relies on the fact that the training is the most computationally-intensive step. However, running only the inference on embedded devices does not solve all problems: a bigger model translates into longer inference times (which sometimes prevent from using the device in real-time applications), and increased energy consumption, \textit{i.e.} reduced battery autonomy. 
If some works manage to embed neural networks intelligently, (see \cite{zhang_towards_2020}, for instance), having a heavy network with many parameters is always a hurdle to efficient embedding. From this point of view, traditional Machine Learning methods, with their reduced requirements, are better suited to be used with embedded systems. 


We will show that it is possible to get the best of both worlds. By using Convolutional Neural Networks instead of the heavier Recurrent Neural Networks we create a deep model that has a small number of parameters and a good classification performance, on two realistic datasets. Also, we show how to expand to CNNs one of the principal quality of RNNs: the ability to use segments of any lengths.

Our contributions are the following:
\begin{itemize}
    \item We reduce the size of a network by a factor of 4 by changing its pooling layer
    \item We demonstrate how to use global pooling methods to allow a convolutional neural network to use segments of different lengths 
    \item We show that the current padding method used to get to segments with equal lengths, the zero-padding, impairs significantly the learning process, and we introduce padding by wrapping, which improves the performance
\end{itemize}

The rest of the paper is organized as follows: section \ref{section:Related work} states the problem and reviews related works. Section \ref{section:Proposed improvements} describes the two improvements we propose, before section \ref{section:Experiments} explains how each of these choices are justified experimentally.

\section{Problem Statement and Related Work}\label{section:Related work}
We first introduce some preliminary definitions (\ref{subsection: Preliminary definitions}) before the statement of the  transport mode detection problem (\ref{subsection:Transport mode detection as a classification problem}). Then we review the related work (\ref{subsection:Related work}). 

\subsection{Preliminary definitions}
\label{subsection: Preliminary definitions}

We will focus on Transport Mode Detection, a problem aiming to guess the transport mode from incoming temporal data. The database consists of a series of \textit{trajectories} (each trajectory being a series of measurements points, such as $(lat, long)$ for GPS signals, or $Acc_x, Acc_y, Acc_z$ (for accelerometer data). Each trajectory has to be divided into \textit{trips}: series of points that are likely recorded in one go (we chose GPS points such that two consecutive points are distant by less than 20 minutes, as in \cite{zheng_learning_2008}). Each trip is made of one or several \textit{triplegs}: series of points sharing a single transport mode. This setting corresponds to a more simple version of the problem, where we know triplegs to have only one mode. In real-life applications, we could consider applying segmentation algorithms like in \cite{zhu_learning_2016, dabiri_semi-supervised_2018, zheng_learning_2008}.

The triplegs might still have different lengths, but a model sometimes requires all inputs to have the same shape. When this is the case, we cut triplegs into fixed-shape \textit{segments}. When we use a model that can deal with arbitrary-sized inputs, segments are equal to triplegs. Thus, a model only classifies segments.

\subsection{Transport mode detection as a classification problem}
\label{subsection:Transport mode detection as a classification problem}

Our goal is to use some labeled data in order to learn to assign a transport mode to unknown data. Using Machine Learning vocabulary, this is a classification problem: we use a certain amount of labelled samples to train a classifier that will predict the transport mode used during the recording of an unseen segment:
\begin{multline*}
 class(segment_i)  \in \{walk, \, bike, \, bus, \, \\car \, \& \, taxi, \, train, \, subway\} 
\end{multline*}


Some additional preprocessing (like computing speed and acceleration) is applied so that the data can be fed to a machine learning model.  In order to train and test properly the model, the dataset will be split into train, validation and test sets. A fraction of the data with the associated labels will be used to train the machine learning model to predict the class of an unseen segment. This is the \textit{train set}. 
We usually have a wide choice when selecting preprocessing functions. In addition, machine learning models generally involve several hyperparameters, \textit{e.g.}, number of filters, or number of layers, that need to be chosen. These hyperparameters are optimized and chosen  in evaluating the variants of the machine learning models on previously unseen segments, the \textit{validation set}. 
Once we have chosen every possible parameter using the validation set, we use the last part of the dataset (\textit{test set}) to evaluate the generalisation skill of the learned model against the state of the art. The following algorithm shows how we use the three datasets:

\begin{algorithmic} 
\REQUIRE Three datasets $X_{train}$, $X_{val}$, $X_{test}$, for training, validation, and testing; a list of hyperparameters sets $L_h$ to evaluate.
\STATE $best\_score \gets 0$
\STATE $best\_hyperparameters \gets  \emptyset$
\FOR{$h$ in $L_h$}
    \STATE Create a deep learning model with hyperparameters $h$
    \STATE Train the model on $X_{train}$
    \STATE Evaluate the model on $X_{val}$, measure the F1-score $score_{val}$
    \IF{$best\_score < score_{val}$}
        \STATE $best\_score \gets score_{val}$
        \STATE $best\_hyperparameters \gets h$
    \ENDIF
    \ENDFOR
\STATE Create a deep learning model with hyperparameters $best\_hyperparameters$
\STATE Train the model on $X_{train}$
\STATE Evaluate the model on $X_{test}$, measure the F1-score $score_{test}$
\RETURN $score_{test}$ for comparison with the state-of-the-art
\end{algorithmic}

\subsection{Related work}
\label{subsection:Related work}

Current research on the subject features two different approaches: traditional machine-learning approaches using handcrafted features and deep learning-based approaches. However, both approaches share a common structure as follows: 
\begin{itemize}
    \item \textit{data cleaning}: When dealing with signals that are are known for being noisy (such as GPS signals \cite{noauthor_gps.gov:_nodate}), most research works use Kalman filters \cite{etemad_transportation_2018}, Savitzky-Golay filters \cite{dabiri_inferring_2018}, or outlier detection thanks to clustering techniques \cite{etemad_transportation_2018} to clean the data. Some approaches also remove the trajectory from the dataset if its speed or acceleration is above a realistic threshold given the class of the trajectory (for instance, the maximum speed and acceleration for the 'bus' class are set to $120~km/h$ and $2~m/s^2$ respectively in \cite{dabiri_inferring_2018})
    \item \textit{point-level feature computation}. When dealing with GPS data, researchers often convert the \textit{(lat,~long)}  into more significant values. Those features are computed at each timestamp. Speed and acceleration are used universally. Other point-level features include distance \cite{zheng_learning_2008}\cite{zheng_understanding_2010}, jerk (the time derivative of acceleration \cite{etemad_transportation_2018}), or delta-bearing (the angle difference at each time step) \cite{etemad_transportation_2018}. 
    \item \textit{segmentation} in single-mode segments. This step is often skipped, most research works prefer using the ground truths to be sure to work on segments containing one transport mode. Those who segment using the data typically rely on walk detection (a small walk is usually necessary between two modes, \cite{zheng_learning_2008}) or the PELT algorithm \cite{killick_optimal_2012, dabiri_semi-supervised_2018}. 
    \item For classical Machine-learning approaches, \textit{trajectory-level feature computation} takes place to get back to a fixed-size feature vector, usable by machine learning models. This treatment often involves computing the mean, standard deviation, minimum, and maximum of each point-level feature \cite{xiao_identifying_2017, etemad_transportation_2018, zheng_understanding_2010}. Other used operations are the computation of percentiles \cite{etemad_transportation_2018, xiao_identifying_2017},  frequency energy bands \cite{maenpaa_travel_2017}, or autocorrelation coefficients \cite{xiao_identifying_2017}. 
    Some trajectory-level features do not rely on a specific point-level feature, such as the stop rate and direction change rate \cite{zheng_learning_2008}, or the closeness to train and bus stations (obtained from other sources, such as Open Street Map \cite{rodriguez-echeverria_methodology_2017}, or Baidou Map \cite{zhu_learning_2016}).
    \item \textit{classification}. For classical Machine-learning approaches, the state of the art is Random Forests \cite{etemad_transportation_2018, rezaie_knowledge_2018} and SVM \cite{zheng_geolife_2019, asgari_transport_2018}. Other classifiers include NaiveBayes \cite{maenpaa_travel_2017}, MultiLayer Perceptrons \cite{maenpaa_travel_2017, rezaie_knowledge_2018}, KNN \cite{xiao_identifying_2017}, HMM \cite{nitsche_supporting_2014}, or even rules \cite{shen_method_2015} and fuzzy rules \cite{sauerlander-biebl_evaluation_2017}. 
\end{itemize}

The 'traditional' Machine Learning approaches use handcrafted features (\textit{e.g.} average speed, maximum acceleration), before using a Machine Learning algorithm (SVM, Random Forest, \textit{etc.}). These approaches are the most simple (computation-wise), but they are also less performing than pure deep learning approaches. 

Within deep learning-based approaches, there is a great diversity of neural networks: Multi Layer Perceptron \cite{endo_classifying_2016, zhu_learning_2016}, Convolutional Neural Networks \cite{wang_detecting_2017, zhang_classifying_2019, dabiri_inferring_2018, vassilev_reconnaissance_2019, dabiri_semi-supervised_2018}, or LSTMs \cite{liu_end--end_2017, yu_travel_2019} (a specific kind of Recurrent Neural Networks).
Some (\cite{wang_detecting_2017, endo_classifying_2016, zhu_learning_2016}) use autoencoders to extract features from trajectories but, curiously, only one \cite{dabiri_semi-supervised_2018} makes use of the unlabeled data in the dataset. This fact is rather surprising, given that unlabeled GPS data is relatively cheap to obtain, contrary to labeled data.

The neural networks that seem to be the best (we will see in section \ref{subsection:Comparison with the state of the art} that comparisons are not straightforward) are the LSTMs, followed by CNNs. However, Recurrent Neural Networks, (including LSTM) rely on matrix multiplications to compute the features for the next layer. This operation is costly in terms of memory, as the weight matrices usually have many parameters. To give an example, the LSTM from \cite{liu_end--end_2017} uses 8 million weights, and 3.2 billion floating-point operations for a single inference, which represents heavy requirements. Contrary to Recurrent Neural Networks, Convolutional Neural Networks rely on convolution operations, which can be extremely cheap, while still retaining good performance levels.

\section{Proposed improvements}\label{section:Proposed improvements}
We present the two main improvements we use to be able to obtain smaller networks with better performance: the choice of an effective pooling layer for convolutional neural networks (section\ref{subsection:pooling}), and the padding of shorter segments (section \ref{subsection:padding}).

\subsection{Pooling operation}\label{subsection:pooling}
\begin{figure}
\centering
\includegraphics[width=3in]{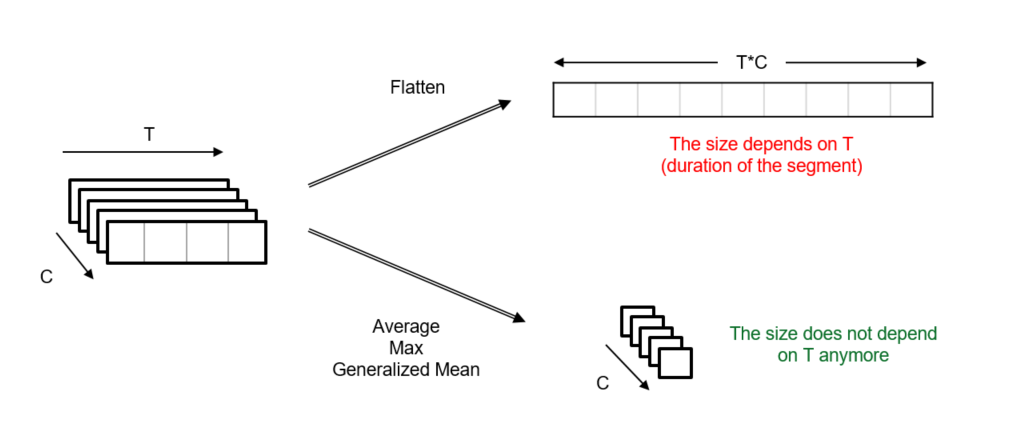}
\centering
\caption{Output sizes of different operations replacing the flatten step. We notice the size of the output of the flatten step depends on both the size of its input along the time dimension (T) and along the channel dimension (C)}
\label{figure:duration_dep}
\centering
\end{figure}

Our convolution layers work with 2-dimensional feature maps (with a \textit{channel} axis and a \textit{time} axis), while the fully-connected layers use 1-dimensional feature vectors. In order to get to a 1-dimensional vector, the most common is to use a flatten operation, a block that simply reshapes the feature map into a vector. However, this operation has a major downside: if the input feature map has $T$ temporal steps and $C$ channels, the next fully-connected will accept features with $T*C$ scalars. As the shape of a fully connected layer cannot change during the training (nor can the number of channels of the previous convolutional layer), the number of time steps $T$ must remain fixed during the training. This means a convolutional network that uses a flatten cannot deal with arbitrary-shaped inputs. By replacing the flatten step with a \textit{global pooling} operation \cite{wang_two-stream_2018}, we obtain a vector which is as long as the input tensor had feature maps, no matter how many time points the input had (see fig. \ref{figure:duration_dep}). This allows to reduce the number of parameters of the network (as the vector is shorter, the next fully-connected layer is much smaller), and to use segments that have arbitrary lengths. 

Several variants of global pooling are possible. If $X_{t,c}$ is the two-dimensional matrix at the end of the last convolution layer ($t$ being the index along the 'time' dimension, while $c$ is the index along the 'channel' dimension), there are several ways to use global pooling. One could use an average over time ($Y_c~=~\frac{1}{T}\sum_{t=0}^{T-1}{X_{t,c}}$) or a maximum ($Y_c~=~max_{t \in 0..T-1}~{(X_{t,c})}$), but a more general option would be to use the generalized mean \cite{tolias_particular_2015}: 

$Y_c = (\frac{1}{T}  \sum_{t=0}^{T-1}{(X_{t,c})^{\alpha_c}})^{1/\alpha_c}$. 

This expression uses one parameter $ \alpha{_c} > 0$ for each channel $c$, which are learnt by gradient descent like any of the other weights. When $\alpha_c=1$, the expression is equal to the arithmetic average. When $\alpha_c\rightarrow+\infty$, the generalized mean converges towards the maximum of the $(X_{t,c})_{t \in [1..T]}$. In order to avoid numerical instability, a small term\footnote{the lowest value we could use was $5.10^{-5}$, which seems quite high} was added to the input tensor $X$. In practice, the values of $\alpha_c$ are initialized following $\mathcal{N}(5,1)$ because we do not want $\alpha_c$ to be negative before the learning even begins (if $\alpha_c<0$, the expression is equivalent to using a generalization of the harmonic mean, which is close to 0 when one of the features $X_{c,t}$ is close to zero, which makes the associated channel useless). The value of all the $\alpha_c$ seem to converge between 0 and 5 during the training process.

\subsection{Padding}\label{subsection:padding}
Padding is the action of filling a short segment with well-chosen values so that the segment reaches a desired shape. This action can be required in two occasions: 
\begin{itemize}
    \item When the model requires a fixed-size input, we cut the triplegs that are too long, which inevitably generates segments that are shorter than the limit. We pad those segments so that they have the same lengths as the others. 
    \item When the model can deal with segments of arbitrary shape, one problem arises: to accelerate the training, the common practice is to parallelize the computation and to submit to prediction a set of 64 or 128 samples (a batch). This requires to put all segments into a single tensor, which is impossible when the segments have different lengths. One could simulate the batch computation by sending each segment one after the other and computing the weight update once after a certain number of segments are processed, but doing so would increase greatly the training times. This is why we pad all the short segments so that they reach the length of the longest segment in the batch. 
\end{itemize}

\begin{figure}
\centering
\includegraphics[width=3in]{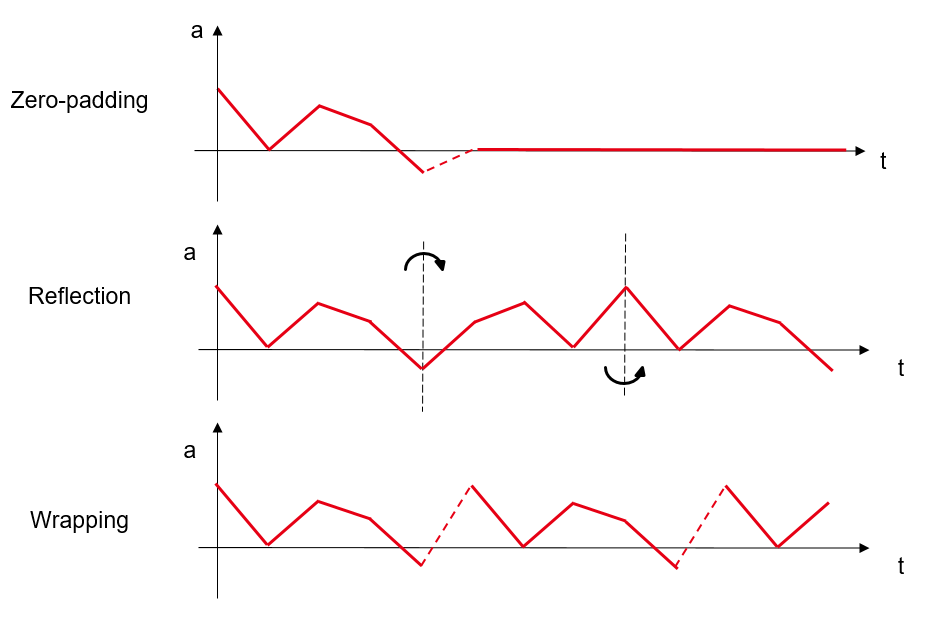}
\centering
\caption{The different kinds of padding. \\Zero-padding simply adds zeros until the maximum length is reached, Reflection reverses a copy of the segment and adds it at the end, while wrapping simply duplicates the segment until the maximum length is reached}
\label{fig:paddings}
\centering
\end{figure}

There are several ways to pad short segments. One could pad using zero-values (like in \cite{dabiri_inferring_2018, dabiri_semi-supervised_2018}), but this disturbs the learning process (see section \ref{subsection:results_padding}). Instead, we pad using the data from the input segment itself (see fig. \ref{fig:paddings}). We tried padding with a reflection of the segment (adding a reversed copy of the segment after the original), or simply repeating the segment until the maximum length is reached. 

We show in section \ref{subsection:results_padding} that padding with zero-values (like in \cite{dabiri_inferring_2018, dabiri_semi-supervised_2018}) is detrimental to the learning process. This is why we chose to pad by wrapping the segment around or by using a symmetric of the segment.



\section{Experiments}\label{section:Experiments}
We begin by giving details about the databases we used (sections \ref{subsection:The GeoLife database}, \ref{subsec:SHL_dataset}). For each dataset, we present the database in itself, along with the preprocessing steps, and the baseline architectures we chose to use. In sections \ref{subsection:results_pooling} and \ref{subsection:results_padding}, we will alter these architectures by changing the pooling and padding (respectively), in order to justify the choices presented back in section \ref{section:Proposed improvements}, before we providing the results of our approach against the state of the art (section \ref{subsection:Comparison with the state of the art}).

\subsection{The GeoLife database}\label{subsection:The GeoLife database}
The GeoLife database \cite{zheng_geolife_2019, zheng_mining_2009, zheng_understanding_2008} was collected between 2007 and 2012. The GPS signals of 180 participants living in five different cities of China were recorded during their commutes, in order to study their behaviours when travelling. Unfortunately, the need for labeled data did not appear immediately, and only one tenth of the trajectories of the database is labeled (in the subsequent, we will only refer to the labeled data, unless otherwise specified). The dataset is an ensemble of trajectories, each trajectory being a series of $(latitude, longitude, timestamp)$ points. Each labeled trajectory has one or more transport modes, and each change between modes has an associated timestamp, so that each point can be attributed a label. The transport modes (classes) in the dataset are: \itshape walk, bike, car, taxi, bus, train, subway\upshape. An overview of the dataset is available in table \ref{table:datasets_overview}.  
We follow the recommendations of the GeoLife user guide \cite{zheng_geolife_2019}, merging together the classes 'taxi' and 'car'. 
One important thing to note is that the data points are not sampled at the same rate: some trajectories have a sampling rate of 1 or 2 Hz, while some others can go down to 0.02 Hz.

\begin{table}
\renewcommand{\arraystretch}{1.3}
\centering
\begin{tabular}{|c|c|c|}
\hline
dataset & GeoLife \cite{zheng_geolife_2019} & SHL 2018 \cite{wang_summary_2018} \\ \hline
number of users  & 69   & 1\\ \hline
total duration  & 5,000 h  & 216 h\\ \hline
total trajectories length & 116,000 km & 1,600 km \\ \hline
\thead{average interval\\between two data points} & 7s & 0.01s\\ \hline
\thead{ median interval\\between two data points} & 2s & 0.01s \\ \hline
\thead{ Total number\\of data point\\in the database}  & $2.5 \times 10^6$ & $9.6 \times 10^7$ \\ \hline
sensors used & GPS & Accelerometer \\ \hline
classes & \thead{Walk, Bike, Bus, \\ Car \& Taxi, Subway,\\ Train} &\thead{Still, Walk, Bike, \\Run, Bus, Car,\\ Subway, Train} \\ \hline
\end{tabular}
\caption{An overview of the labeled data in the GeoLife and SHL 2018 datasets}
\label{table:datasets_overview}
\end{table}

As we explained in section \ref{subsection:Transport mode detection as a classification problem}, we split the triplegs into three sets, training, validation, and testing. For the GeoLife dataset, 64\% of the triplegs go in the train set, 16\% in the validation set, and 20\% in the test set. With this dataset, we chose the hyperparameters using Random Search \cite{bergstra_random_2012}. We fix a possible range of hyperparameters using usual values found in the literature, and train several models, each model using a series of hyperparameters at random. We then selected the hyperparameter which improved the performance on average, by looking at the median and quartiles of the validation F1 (results not shown). We also use the validation set for early stopping in each of the training process, during the random search or to produce the results in the present work.

\subsubsection{Preprocessing}\label{subsection:Data preprocessing}
We begin by computing the speed and acceleration of each point. This way, our data is not dependent on the precise location of the trajectory. We remove the data points which acceleration or speed are deemed unrealistic given the annotated transport mode (we reused the values from \cite{dabiri_inferring_2018}). We considered adding the bearing \cite{dabiri_inferring_2018}, but it turned out using this feature did not increase the performance of our model. As the sampling rate is very irregular, we interpolate linearly our data points (T~=~2s), so that a difference between two consecutive points always has the same meaning.

We do not apply any cleaning or filtering, for we found these to be unnecessary.

\subsubsection{Data Preparation and Splitting} \label{subsection:The separation of segments}
Etemad \cite{etemad_transportation_2018} showed that the way the segments are split between the different sets (training, validation, test) can have a huge influence on performance: when a tripleg is split into several segments and the segments of a single trajectory can go in both the training and the test set, the trained model will be likely to have seen parts of all trajectories in the dataset, which will cause it to overfit. 

In his experiments, Etemad found the F1 score can be vary by 20 percentage points between a training scenario in which the users are correctly split (71 \%), compared to a scenario in which the fragments of a given trajectory can go in different sets (91 \%). This result is not surprising, as mobility trajectories have a high degree of regularity  \cite{gonzalez_understanding_2008, song_limits_2010, huang_exploring_2020}. Ideally, we should train a prediction model using the trajectories from a set of users and test the generalization skills of the trained model on those of unseen users. This means that we need first to assign the trajectories of each user  to one set among training, validation or test (as in \cite{etemad_transportation_2018, endo_classifying_2016, dabiri_inferring_2018, dabiri_semi-supervised_2018}). This corresponds to the most realistic setting, where an algorithm predicts the transport mode of unseen users.

But in practice, with the GeoLife dataset, this method leads to extremely imbalanced sets, as users have different habits when it comes to transportation. In some cases (depending on the seed used to initialize the splitting process), splitting the dataset by users can even produce validation or test sets that completely lack one class. To show this, we realized 200 separations with different seeds initializing the random process, and looked at the effect it would have on the final distribution (results not shown). If the median is centered around the correct distribution, the quartiles show a high variance.  In the validation and test set, the third quartile is at least twice higher than the first quartile. This distribution variance is a problem for comparison, because the performance of imbalanced models will strongly depend on the distribution, even when using measures like the F1 score.

This is why we only split the sets by tripleg: we first separate triplegs between train (64\% of the trajectories), validation (16\%), and test (20\%), before segmenting the triplegs into segments. This method is less realistic than splitting the sets by users, but it allows to produce sets with similar distributions consistently. The following pseudo-code explains how we split by triplegs:

\begin{algorithmic} 
\REQUIRE A list $L_{users}$ of users, each user being a list of trips
\STATE Create a list $L_{trips}$ of trips by merging all users
\STATE Randomly split the list $L_{trips}$ into three lists $\{L_{trips}^{train}, L_{trips}^{val}, L_{trips}^{test}\}$
\FOR{$s$ in $\{train, val, test\}$}
    \STATE $X_s \gets \emptyset$
    \FOR{each trip $t$ in $L_{trips}^s$}
        \STATE Split the trip $t$ into triplegs and,
        \STATE Add the triplegs to the set:
        \STATE $X_s \gets X_s \bigcup t_{split}$ 
    \ENDFOR
\ENDFOR
\RETURN the three tripleg sets $X_{train}, X_{val}, X_{test}$ 
\end{algorithmic}

\begin{table}[!t]
\renewcommand{\arraystretch}{1.3}
\centering
\scriptsize
\centering
  \subfloat[GeoLife]{ \begin{tabular}{|c|c|c|c|c|c|c|}
  \hline
             & walk & bike & car \& taxi & bus  & subway & train \\ \hline
total        & 4517 & 1731 & 1459        & 2129 & 632    & 200 \\ \hline
train        & 2890 & 1108 & 934         & 1362 & 405    & 128 \\ \hline
validation   & 723  & 277  & 233         & 341  & 101    & 32 \\ \hline
test         & 904  & 346  & 292         & 426  & 126    & 40 \\ \hline

\end{tabular}
  \label{table:distribution_trajectories_GeoLife}}
\hfil
\subfloat[SHL]{
\begin{tabular}{|c|c|c|c|c|c|c|c|c|}
  \hline
            & still & walk & run & bike & car  & bus  & subway & train \\ \hline
total       & 2302  & 2190 & 686 & 2101 & 2475 & 2083 & 2520 & 1953  \\ \hline
train       & 1899  & 1630 & 506 & 1716 & 2141 & 1663 & 1973 & 1472    \\ \hline
validation  & 403   & 560  & 180 & 385  & 334  & 420  & 547 & 481\\ \hline

\end{tabular}
  \label{table:distribution_trajectories_SHL}}
\caption{The number of triplegs in each subset the GeoLife (a) and SHL 2018 (b) datasets. Note that the GeoLife dataset is more balanced than the SHL dataset}

\label{table:distribution_trajectories}
\end{table}

\subsubsection{Baseline architecture}\label{subsubsection:GL_architecture}

\begin{figure*}
    \centering
    \includegraphics[width=7in]{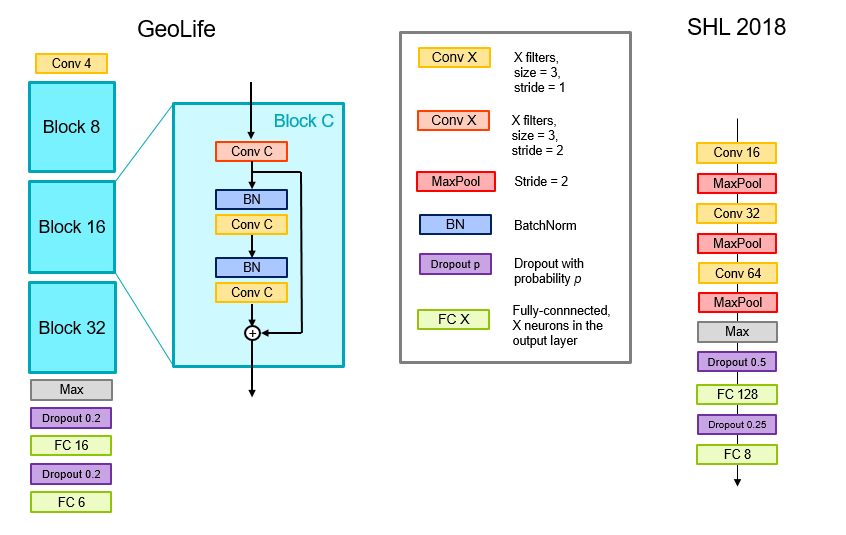}
    \centering
    \caption{The baseline architectures of both networks}
    \label{figure:final_architectures}
    \centering
\end{figure*}

\begin{table}
    \renewcommand{\arraystretch}{1.5}
    \centering
    \begin{tabular}{c c c}
        \hline
        Parameter                           & value (GeoLife)      & value (SHL)     \\ \hline
        learning rate                       &  $ 10^{-2}$          & $ 10^{-3}$      \\ 
        regularization parameter            & $ 3.10^{-3}$         & $ 10^{-3}$      \\ 
        batch size                          & 128                  & 64              \\ 
        non-linearity                       & ReLU                 & ReLU            \\ 
        optimizer                           & Adadelta \cite{zeiler_adadelta:_2012}  & Adam            \\
        max number of epochs                & 2000                 & 50\\    
        patience                            & 100                  & / \\ \hline
    \end{tabular}
    \caption{The chosen hyperparameters for the training of both models}
    \label{table:hyperparameters}
\end{table}

The GeoLife CNN uses the $(speed, acc)$ features to classify a segment. Its architecture is inspired by ResNet \cite{he_deep_2016} and is given in fig. \ref{figure:final_architectures}. As explained in the section \ref{section:Related work}, the use of convolutions allow to obtain a particularly efficient architecture compared to the recurrent models, especially in terms of memory footprint (see table \ref{table:results}).

To train our neural network, we use a weighted version of the cross-entropy loss: the loss of every segment is given a weight that is inversely proportional to the number of elements in this segment's class on the training set (which is proportional to the proportion on the whole dataset, see section \ref{subsection:The separation of segments}). The goal of this procedure is to compensate for the class imbalance in the dataset. We also apply early stopping: we stop the training process when the loss on the validation set did not increase for more than a fixed number of epochs (chosen to be 100, see table \ref{table:hyperparameters}), and we keep the model which minimized the validation loss for testing. Usually, this minimum loss is reached between epoch 100 and 600.

\subsection{The SHL 2018 database}\label{subsec:SHL_dataset}

The SHL 2018 dataset consists in a series of 16,310 consecutive annotated recordings of embedded sensors. Each recording is 60-seconds long, and contains data from 7 sensors, and 20 channels: three axes (\textit{x, y, z}) for the accelerometer, magnetometer, gravity, linear acceleration (acceleration minus gravity), and gyrometer; one orientation quaternion (\textit{x, y, z, w}), and a recording of the barometric pressure. Each signal was recorded at $100 Hz$, so that one sample to classify is a set of 20 vectors of size $60 \times 100 = 6000$ points. There are 8 classes available: Still, Walk, Run, Bike, Car, Bus, Train, Subway. 

The aim here is to check the general nature of our method using a different dataset, in particular the finding that a well-chosen padding and global average pooling help improve the performance and reduce the memory footprint of the networks. 


\subsubsection{Preprocessing}\label{subsubsec:SHL_preprocessing}

The SHL dataset is particular, because the triplegs of the dataset all have the same length. Padding is unnecessary when all segments are the same lengths. Hence, we choose to work with a degraded dataset: for each segment, we randomly keep between $10$ and $100 \%$ of the data (uniform probability), and we pad the rest by wrapping the segments (this choice is justified experimentally in section \ref{subsection:padding}). When we evaluate a result, we generate a new set (thereby choosing a new proportion for padding for each segment) each time we train a new network (5 times in total). This is why the standard deviations in tables \ref{table:results_pooling} and \ref{table:results_padding} are so high: they account for network performance and dataset variation. 

As we do not aim to overcome the state of the art with this set, we focus on the single bet sensor: the accelerometer \cite{ito_application_2018, wang_summary_2018}. We compute the norm of the accelerometer $Acc_{norm} = \sqrt{Acc_x^2+Acc_y^2+Acc_z^2}$ as our input signal, this will be our only preprocessing (apart from the degradation of the dataset mentioned earlier). As the accelerometer signal is not particularly noisy, we do not apply any cleaning. We do not even remove the trajectories with unlikely values, because the creators of the dataset already double-checked their annotations with videos they recorded.

In order to split the training set from the validation set, we make sure that we split the data not to have fragments of a single trip go into both the train and the val set \cite{widhalm_top_2018}. The segments are sorted chronologically, and we take the first 20\% of the segments for the validation set and the last 80\% for the training set (doing so ensure both sets are approximately balanced, as the end of the dataset is severely imbalanced). 
We use the validation set for comparisons, but, contrary to the algorithm in section \ref{subsection:Transport mode detection as a classification problem}
showed, but we do not use the test set for this dataset. 

\subsubsection{Baseline architecture}
Our baseline architecture, along with its parameters, come from \cite{ito_application_2018}. The CNN uses a segment with input size $6,000 \times 1$ (as all our segments were padded back to get 6,000 points), to return a prediction among the 8 classes (Still, Walk, Run, Bike, Car, Bus, Train, Subway). Figure \ref{figure:final_architectures} shows the architecture, while table \ref{table:hyperparameters} shows the selected hyperparameters. 
Contrary to the GeoLife model, we only train the SHL model for 50 epochs, after which we evaluate it and return the validation score (similarly to \cite{ito_application_2018}).

\subsection{Experimental setup}

In order to measure the performance of our models, we use the F1-score. This measure derives from the accuracy, except that it penalizes heavily the models which have imbalanced predictions. In the case of Transport Mode Detection, this is almost always the case. To make sure of this, one can look at the distributions in the SHL and GeoLife datasets (tables \ref{table:distribution_trajectories_GeoLife} and \ref{table:distribution_trajectories_SHL}, respectively).

To evaluate the complexity of each models, we display their number of parameter, along with the number of operations for a single inference (forward pass). As an illustration, we did include training and evaluation times, even though these results depend heavily of the device. For instance, GPUs are more optimized for heavily parallel operations, such as convolutions. The number of parameters and operations, however, is independent of the implementation, and provides an objective measure for comparison. 
We trained the models on a server with a Nvidia tesla V100 GPU (32 Gb of memory), cuda version was 10.2, and a 40-core Intel Xeon Gold 6230 CPU @ 2.10GHz with 190 Gb of RAM. The evaluation times were measured on a CPU, a 4-core Intel i7-7820 @ 2.90 GHz with 32 Gb of RAM. However, those running times are not absolute measures: some devices are better optimized for different types of operations. The only objective measurements are the number of parameters and operations, which do not depend on the device

For each result (F1-score, training times), we repeat the training and devaluation process 5 times, changing the seed each time. We display the average $\pm$ standard deviation.

\subsection{Pooling operation}\label{subsection:results_pooling}

\begin{table*}
\footnotesize
\renewcommand\theadfont{\footnotesize}
\renewcommand{\arraystretch}{1.8}
\centering
\begin{tabular}{ |c|c|c|c|c|c|c|c|}
\hline
&   Pooling       & \thead{Validation\\F1-score}     & \thead{number of\\parameters}  & \thead{operations\\(FLOPs)} & training time (min) & \thead{epochs to \\convergence} &  inference time ($ms$) \\  \hline
\multirow{4}{*}{ \rotatebox[origin=c]{90}{GeoLife}}    & \thead{Flatten\\(segments of 1,024 points)} 
                             & $ 77.0 \pm 1.6 \%$  &  $7.6 \times 10^4$  &  $9.4\times10^4$  &  $7.8 \pm 0.7$ &  $113 \pm 17$  & $1.92 \pm 0.13$\\ \cline{2-8}
 &\thead{Generalized Mean}   & $ 80.9 \pm 1.0 \%$  &  $1.1 \times 10^4$  &  $3.3\times10^4$  & $43.9 \pm 7.3$ &  $538 \pm 114$ & $1.94 \pm 0.04$  \\ \cline{2-8}
 & Average                   & $ 80.2 \pm 1.3 \%$  &  $1.1 \times 10^4$  &  $3.3\times10^4$  & $78.7 \pm34.6$ & $1161 \pm 571$ & $1.81 \pm 0.04$  \\ \cline{2-8}
 & Maximum                   & $ 80.3 \pm 1.6 \%$  &  $1.1 \times 10^4$  &  $3.3\times10^4$  & $16.8 \pm 2.7$ &  $262 \pm  66$ & $1.85 \pm 0.06$ \\ \hline
\multirow{4}{*}{ \rotatebox[origin=c]{90}{SHL}}    &  \thead{Flatten\\(segments of 6,000 points)}   
                             & $ 65.2 \pm 3.2 \%$  &  $6.1 \times 10^6$  &  $4.2\times10^7$  & $1.20 \pm 0.00$ &      $50$     & $4.74 \pm 0.02$  \\ \cline{2-8}
 &\thead{Generalized Mean}   & $ 67.5 \pm 1.7 \%$  &  $1.7 \times 10^4$  &  $3.0\times10^7$  & $1.15 \pm 0.09$ &      $50$     & $3.60 \pm 0.43$ \\ \cline{2-8}
 & Average                   & $ 59.8 \pm 2.6 \%$  &  $1.7 \times 10^4$  &  $3.0\times10^7$  & $0.97 \pm 0.00$ &      $50$     & $2.60 \pm 0.02$ \\ \cline{2-8}
 & Maximum                   & $ 68.7 \pm 1.1 \%$  &  $1.7 \times 10^4$  &  $3.0\times10^7$  & $1.03 \pm 0.06$ &      $50$     & $2.68 \pm 0.04$ \\ \hline
\end{tabular}
\caption{The effectiveness of each kind of pooling, in terms of performance, computational resources, and training and inference time. For each result, we display the average and the standard deviation, over 5 runs}
\label{table:results_pooling}
\end{table*}

We compared three alternatives to the Flatten, namely Maximum, Mean, and Generalized Mean. As table \ref{table:results_pooling} shows, on the GeoLife dataset, only the flatten step is worse than the rest, in terms of performance, computational requirements, and training and testing time.  
On the SHL dataset, the global max-pooling is significantly better than both the flatten step and the global average. Surprisingly, the Average pooling has a worse performance than the flatten step. The max-pooling, however, is better performance-wise and complexity-wise. Finally, it should be noted that, for both datasets, the general mean is not significantly higher than the rest, even though this operation encompasses the simple average and maximum poolings. We did not help the network by providing it with a more general expression.

As for the running times, table \ref{table:results_pooling} shows us that the use of alternatives to the flatten step make the training times \textit{longer}. The convergence is slower with these architectures, which cancels the speed gain of these architectures. As most applications rely on training the model offline (on a computer), the training time of a model is not a major concern. More interesting is the inference time: we can see that the global pooling operations are up to $X \%$ faster than the flatten step they replace.

\subsection{Padding}\label{subsection:results_padding}

Using the baselines architectures (fig. \ref{figure:final_architectures}), we compared the three kinds of padding shown in fig. \ref{fig:paddings}: 
\begin{itemize}
    \item \textit{Zero-padding}, where zeros are added to the shorter segments until they reach the correct length, as in \cite{dabiri_inferring_2018, dabiri_semi-supervised_2018}.
    \item \textit{Wrapping}, where segments are padded using their own data
    \item \textit{Reflection}, which consists in padding the segments using a time-reversed version of the segment itself.
\end{itemize}

As table \ref{table:results_padding} shows, padding with zeros is particularly detrimental to the performance of our model. However, one can wonder which padding is better between wrapping and reflection. Wrapping creates discontinuities in the data, but this is the only artifact it introduces: otherwise, wrapping only uses data from the segment itself. On the other hand, reflection removes some of the meaning of the data (\textit{i.e,} and acceleration becomes a breaking), but it does not introduce any discontinuity. 

The GeoLife model has a particularity: during the random search, we contemplated adding a cleaning step using median or Savitzky-Golay filters, and it turns out these filters did not improve the performance (results not shown). This means the GeoLife network is naturally robust to noise in the data, which is why it is not affected by the discontinuities brought by wrapping the segments around. 
We hypothesize the same reasoning applies for the SHL network. In this case, the direction of time matters much mode than the discontinuities we brought by using wrapping.

One could wonder why zero-padding is worse than the rest. Two mutually exclusive hypotheses could be formulated to explain this phenomenon:
\begin{itemize}
    \item A long series of zeros is interpreted as being meaningful by the model, and disturbs its \textit{predictions}. 
    \item The model notices the long series of zeros in the learning process, and, upon seeing they are uncorrelated with the segment's classes, somehow learns to ignore the end of a segment during the training process, which cause it to miss relevant information
\end{itemize}

\begin{figure*}
\centering
\includegraphics[width=6.3in]{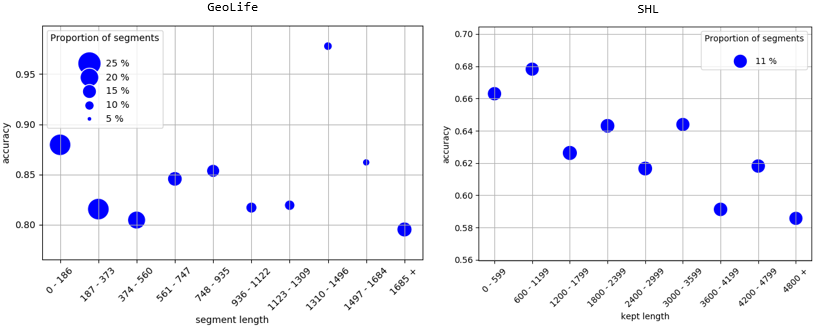}
\centering
\caption{The validation accuracy versus the size of each segment (the shorter the segment, the more zeros it will be padded with). Adding zeros is not particularly detrimental to the classification performance, which means the network learnt to ignore the zeros, missing potentially relevant information. Intervals bins obtained using equidistant separations between 0 and the 90-th percentile}
\label{figure:acc_vs_zeros}
\centering
\end{figure*}

To know which one is true, we compute the accuracy for segments with different lengths (using the fact that the shorter the segment, the more zeros it will be padded with). If the former hypothesis was true, we would see the performance to be correlated negatively with the number of zeros. The performance would be correlated positively with the length of a segment, resulting in a decreasing curve in fig. \ref{figure:acc_vs_zeros}. 

Here, the behaviour depends on the dataset: the GeoLife curve is quite irregular, which means that the model learnt to ignore zeros at the end of segments. For the SHL model however, the performance is clearly negatively correlated with segment length: the model does not know that zeros are meaningless, and tries to interpret them, leading to higher errors when zeros are more present.

\begin{table}
\renewcommand{\arraystretch}{1.5}
\centering
\begin{tabular}{ c|c|c}
\hline
 Padding    &    GeoLife          &       SHL             \\ \hline
 Zero       & $ 77.7 \pm 1.5 \%$  &  $ 64.2 \pm 3.5 \%$   \\ 
 Reflection & $ 80.2 \pm 1.6 \%$  &  $ 63.9 \pm 1.9 \%$   \\ 
 Wrapping   & $ 80.3 \pm 1.6 \%$  &  $ 68.7 \pm 1.1 \%$   \\ \hline
\end{tabular}
\caption{The validation F1-score of each type of padding. Zero-padding is particularly detrimental to the model performance}
\label{table:results_padding}
\end{table}

\subsection{Comparison with the state of the art}\label{subsection:Comparison with the state of the art}

\begin{table*}[t]
\centering
\renewcommand{\arraystretch}{1.5}
\begin{tabular}{|c|c|c|c|c|c|c|c|}
\hline
   \thead{model} & \thead{metric} & \thead{reported\\score}  & \thead{classes} & \thead{are trajectories\\properly split ?} & \thead{are trajectories\\already\\segmented ?} & \thead{Number of\\parameters} & \thead{Number of\\ operations\\ (FLOPs)} \\  \hline\hline
LSTM + embedding \cite{liu_end--end_2017}                  & AUC & 94.5 \%  & 4 &  NM & yes &  \thead{$1.1\times 10^6$  \\($ 100 \times p $)}      &  \thead{   $4.2\times 10^8$ \\($ 10,000 \times o $) }\\ \hline 
Proposed                                      & F1  &  $ 87.1 \pm 1.1 \%$   & 4 & yes & yes &  \thead{$1.1\times 10^4$  \\ $(p)$}                           &     \thead{$3.3\times 10^4 $ \\ $(o)$}\\ \hline\hline
AE + CNN \cite{dabiri_semi-supervised_2018}                 & F1 & 76.4 \%  & 5 &  NA & no  &  \thead{$4.1\times 10^4$ \\($ 4\times p $) }        &  \thead{   $6.4\times 10^6$ \\($ 100 \times o $)}\\ \hline 
 CNN ensemble (7 models)~\cite{dabiri_inferring_2018}       & F1 & 84.0 \%  & 5 & yes & yes &  \thead{$7 \times 2.6 \times 10^6$ \\($ 1,000 \times p $)} & \thead{$ 7 \times 1.7\times 10^7$ \\($ 1,000 \times o $)}\\ \hline
LSTM + Wavelet features \cite{yu_travel_2019}               & F1 & 91.9 \%  & 5 &  NM & yes &  \thead{$8.1\times 10^6$    \\($ 1,000 \times p $)}  &   \thead{  $7.3 \times 10^9$ \\($ 100,000 \times o $) }\\ \hline 
Proposed                                      & F1 & $ 83.9 \pm 1.1 \%$     & 5 & yes & yes &  \thead{$1.1\times 10^4$  \\ $(p)$}                           &     \thead{$3.3\times 10^4 $ \\ $(o)$}\\ \hline\hline
 Random Forests \cite{etemad_transportation_2018}           & F1 & 71   \%  & 6 & yes & yes &   50 trees           & /                    \\  \hline
 
Proposed                                          & F1 & $ 81.8 \pm 0.9 \%$ & 6 & yes & yes &  \thead{$1.1\times 10^4$  \\ $(p)$}                           &     \thead{$3.3\times 10^4 $ \\ $(o)$}\\ \hline\hline
AE + Logistic Regression \cite{endo_classifying_2016}  & accuracy & 67.9 \% & 7 & yes & yes &  \thead{$2.7\times 10^5$ \\($ 10 \times p $) }       & \thead{ $5.2\times 10^5$ \\($ 10 \times o $) }\\  \hline 
Proposed                                         & F1 & $ 74.1 \pm 0.7 \%$  & 7 & yes & yes &  \thead{$1.1\times 10^4$  \\ $(p)$}                           &     \thead{$3.3\times 10^4 $ \\ $(o)$}\\ \hline\hline
 CNN (DenseNet + Attention) \cite{zhang_classifying_2019}   & F1 & 72.0 \% & 14 &  NM & yes &  \thead{$1.3\times 10^5$  \\($ 10 \times p $)}       & \thead{  $7.2\times 10^7$ \\($ 1,000 \times o $)}\\  \hline  

\end{tabular}
\caption{Final results, on the GeoLife test set. We display a performance metric (as provided by the cited works), the estimated size (the number of weights) of each model, and the estimated number of operations required for one classification forward pass. With each of these values, we also display the ratio with the number of parameters ($p$) or operations ($o$) of our model. NM: not mentioned.  NA: not applicable (trajectories are not segmented into triplegs and can have several transport modes)}
\label{table:results}
\end{table*}

\begin{figure}
\centering
\includegraphics[width=3in]{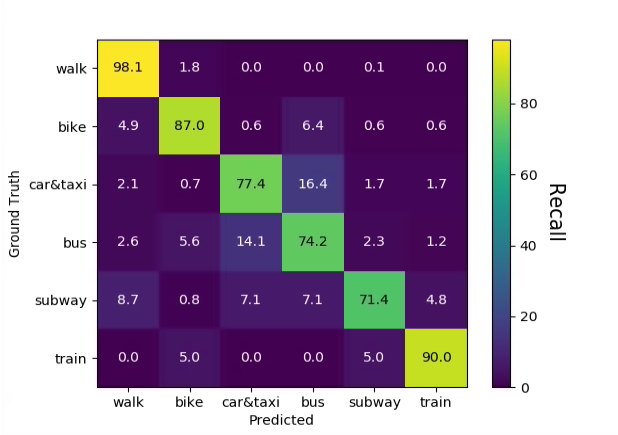}
\centering
\caption{The confusion matrix of the GeoLife model, on the test set. We can see merging together the "car \& taxi" and "bus" classes will improve the performance considerably.}
\label{figure:confusion}
\centering
\end{figure}

We show the improvements one can achieve on the GeoLife dataset (we do not compare the SHL model against the state of the art because it relies on using a degraded dataset). When evaluating a model, a close attention must be paid to all the classes every publication worked with: in all our previous experiments, we strictly followed the recommendations of the GeoLife user guide \cite{zheng_geolife_2019}, which advises to merge together the classes 'taxi' and 'car', for the number of segments in the former is deemed too small.

To know how a change in the categories employed would affect the results, one could look at the confusion matrix (fig. \ref{figure:confusion}). For instance, if two modes are frequently confused with each other, merging them would cause the model's F1 to increase. This applies to the merger of \{car \& taxi\} with bus and, to a lesser extent, to the merger of the classes train with subway. 
A similar reasoning could be made about the removing of the two rail modes (leaving only four modes, \{walk, bike, car, bus\}, as \cite{liu_end--end_2017} do): not only are these modes no longer confused with each other, but they are not even confused with car. As one could expect, the number of classes employed correlates negatively to a model's performance. 

Noticeably, \cite{endo_classifying_2016} choose not to merge any classes, while \cite{zhang_classifying_2019} add another difficulty: not only they do not merge transport modes, but they also try to predict whether the trajectory was "fast" or "slow" (above a threshold depending on the transport mode \textit{eg} 25 m/s for the car). A trajectory is correctly classified only if both the transport mode (7 classes) and the speed category (2 classes per transport mode) are correct. This separation makes sense in their setting, because the representation they use for a trajectory (binary heatmaps) prevents the speed information from being directly transmitted to the deep learning model. Finally, it should be noted that \cite{dabiri_semi-supervised_2018} does work with unsegmented trajectories, which makes the challenge harder.

The number of classes vastly impact a model's performance and the problem is harder with more classes. Nevertheless, to enable comparison with the state of the art despite the class difference, we retrain five models on each class combination, and evaluate them on the test set  (in this section, the validation set is only used to find the loss minimum for early stopping). However, the model we chose to train still keeps the hyperparameters and the architecture we found using the 6-class problem (see section~\ref{subsubsection:GL_architecture}). 
The results are in table~\ref{table:results}.  We can see that our approach creates a network which is both small and effective, for our model manages to compete with the state of the art, while reducing the number of parameters by a factor 4 to 1,000, and the number of Floating-point operations (FLOPs) by a factor 10 to 100,000.

For the 4-class problem, even if the measures differ, the performance of our approach seems lower than the reported performance of the corresponding publication, but the required resources are also much lower. Besides, the corresponding work (\cite{liu_end--end_2017}) does not say whether segments from the same trajectory can go in same or different sets.
For the 5-class problem, our approach is better or equal to existing Convolutional neural networks performance-wise. If the residual CNN is worse than the LSTM implemented in \cite{yu_travel_2019}, this publication does not precise how the train, validation, and test set were split. 
On the 6-class problem, our residual CNN is better than the Random Forest presented in \cite{etemad_transportation_2018}. However, even if the author does not leave enough to precisely estimate the memory and computation requirements of their approach, it is likely that our CNN has worse requirements, despite these requirements being quite low for a typical CNN. 
On the 7-class problem, the residual CNN is better than the autoencoder from \cite{endo_classifying_2016}, both performance and constraint-wise. 
We do not look at the 14-class problem because, with our encoding (we compute the speed explicitely), it is equivalent to the 7-class problem. 

\subsubsection*{Memory and computation costs}

Two measures matter when it comes to comparing the efficiency of different neural networks: the number of parameters (which indicates how much memory will be needed to store the network), and the number of operations required for a single inference\footnote{As neural networks are typically trained offline before being used for predictions, we did not look at training costs in this part} (forward pass). Looking at the time needed for an inference (as in \cite{yu_travel_2019}) is a good measure to evaluate if a network can be used in a client-server architecture, but it is device-dependant.  Contrary to computing times, the number of parameters and operations can be computed from the architecture of a network, and allow for clear comparison (table \ref{table:results}).

We computed the number of operations using the code from \cite{sovrasov_sovrasovflops-counterpytorch_2020}. When the input shape may vary (as in \cite{liu_end--end_2017, yu_travel_2019} or the present work), we used the median length of a segment in the dataset: 200 points on the original dataset (for \cite{liu_end--end_2017, yu_travel_2019}), 500 points on the interpolated one (with our proposed model). \textit{Note}: for \cite{liu_end--end_2017}, we chose the network with a hidden layer of 100 elements instead of 300, even if it is slightly worse (94.5~\% AUC instead of 94.6~\%). As we aim to find a good balance between performance (F1 score, AUC, or accuracy) and efficiency, we estimated selecting a network with three times more weights for a .1 percentage point gain was unfair.

\section{Conclusion}\label{section:Conclusion}

Using experiments on two realistic datasets, we showed that using adequate pooling operations could drastically reduce the number of parameters of a network without impairing its performance, which allowed us to obtain networks with the same level as the state of the art, but with 10 to 1,000 times less parameters. 
Also, we showed the global pooling operation allows to get rid of one of the limits of Convolutional Neural Networks: the limitation of input size. 
This limit required to use an operation (the padding operation) which impaired the learning process when used as implemented in the literature. By padding with data from the segment instead of zeros, we could increase the performance. 
 
We consider three directions for future work: we could decide to work with full, unsegmented, trajectories, in order to do both the segmentation and the classification at once. 
Alternatively we could try to solve the open-set problem. Currently, our network tries to classify every segment it sees, even if these segments does not belong in any class from the training set. This means segments from unknown classes (whether they belong to classes we removed from the GeoLife dataset, \textit{e.g.}, boat, motorbike, plane, or to completely novel transport modes which might appear in the future). We could consider adding an extra class, named 'unknown' or 'other' so that our model does not try to assign a defined class to novel segments. 
A last direction for improvement is to use the promising attention-based architectures (\textit{e.g.}, transformer). These models have achieved impressive results on natural language processing and even computer vision, at the cost of a massive increase in complexity. Finding a way to make them usable with a reasonable amount of parameters or operations might be the next hurdle for Transport Mode Detection.

\bibliography{main}

\end{document}